\documentclass[%
 aip,
 amsmath,amssymb,
 reprint,%
]{revtex4-1}

\DeclareUnicodeCharacter{2009}{\,}
\usepackage{graphicx}
\usepackage{dcolumn}
\usepackage{bm}

\usepackage[utf8]{inputenc}
\usepackage[T1]{fontenc}
\usepackage{mathptmx}
\usepackage{etoolbox}

\makeatletter
\def\@email#1#2{%
 \endgroup
 \patchcmd{\titleblock@produce}
  {\frontmatter@RRAPformat}
  {\frontmatter@RRAPformat{\produce@RRAP{*#1\href{mailto:#2}{#2}}}\frontmatter@RRAPformat}
  {}{}
}%
\makeatother
\begin{document}

\preprint{AIP/123-QED}

\title[Determination of Optimal Chain Coupling made by Embedding in D-Wave Quantum Annealer]{Determination of Optimal Chain Coupling made by Embedding in D-Wave Quantum Annealer}
\author{Hayun Park}
 \altaffiliation{Department of Liberal Studies, Kangwon National University, Samcheok, 25913, Republic of
Korea}
\author{Hunpyo Lee}%
\email{hplee@kangwon.ac.kr}
\affiliation{Department of Liberal Studies, Kangwon National University, Samcheok, 25913, Republic of
Korea
}%
\affiliation{Quantum Sub Inc., Samcheok, 25913, Republic of Korea}

\date{\today}

\begin{abstract}
The qubits in a D-wave quantum annealer (D-wave QA) are designed on a Pegasus graph that is
different from structure of a combinatorial optimization problem. This situation requires embedding with
the chains connected by ferromagnetic (FM) coupling $J_c$ between the qubits. Weak and strong
$J_c$ values induce chain breaking and enforcement of chain energy, which reduce the accuracy of
quantum annealing (QA) measurements, respectively. In addition, we confirmed that even though the D-Wave
Ocean package provides a default coupling $J_c^{\text{default}}$, it is not
an optimal coupling $J_c^{\text{optimal}}$ that maximizes the possible correct rate of QA measurements.
In this paper, we present an algorithm how $J_c^{\text{optimal}}$ with the maximum probability $p$ for
observing the possible lowest energy is determined. Finally, we confirm that the extracted
$J_c^{\text{optimal}}$ show much better $p$ than $J_c^{\text{default}}$ in QA measurements of various
parameters of frustrated and fully connected combinatorial optimization problems. The open code is 
available in \textit{https://github.com/HunpyoLee/OptimizeChainStrength}.
 
\end{abstract}

\maketitle

\section{Introduction\label{Introduction}}

The recent progress in quantum technology has led to the emergence of quantum machines. In addition, they
are being developed and built more frequently than before~\cite{Preskill2018}. In particular, a D-Wave
quantum annealer (D-Wave QA) is an example of such a quantum machine~\cite{Johnson2011}. Unlike gate-type
quantum machines based on circuits, quantum annealing (QA) is implemented in the
parameterized Hamiltonian of a transverse-field Ising model containing binary superconducting
qubits~\cite{Kadowaki1998,King2022,Das2008}. The primary advantage of this architecture quantum machine is 
that
qubits are added much easier than gate-type quantum computers with maintaining the accuracy of the
results, due to short quantum coherence time. Thus, the D-Wave QA with rapid increases of qubit capacity
approaches computational speed of classical digit machine. In addition, studying the dynamic behaviors 
observed in real materials is a very interesting research 
topic, as it is not easy to study using numerical simulation methods. Ca$_3$Co$_2$O$_6$ compound is an 
example with unconventional dynamics that has remained a puzzle~\cite{Nekrashevich2022}. Recent QA 
experiment on frustrated spin 
system reported that it can be used as a programmable quantum simulator for both the dynamic and 
equilibrium behaviors observed in Ca$_3$Co$_2$O$_6$ compound~\cite{King20211}. Consequently, this 
quantum machine has been not only employed in combinatorial optimization problems requiring annealing 
process for the possible lowest energy and for the research of Ising model, which exhibits the phase 
transition between unconventional phases at zero temperature, but also used as programmable quantum 
simulator for exploring dynamic behaviors shown in real materials~\cite{King20211,Amin2018, Isakov2016, 
Mazzola2017,Inoue2021, Kairys2020,King2021, Irie2021,Park2022,Ronnow2014,Albash2018,Ronnow20141}.

However, the D-Wave QA cannot technically describe the exact couplings between long-distance qubits, and
it retains qubits designed on the specific architectures such as Pegasus and Kimera graphs that depend on
D-Wave QA hardware types. These situations require physical embedding that the architecture of the
original problem topologically matches that of D-Wave QA~\cite{Lanthaler2021,Konz2021}. The auxiliary
chains with ferromagnetic (FM) coupling $J_c$ between several qubits, represented as one variable
in the architecture of the original problem, are required. Weak and strong $J_c$ induce chain
breaking and reinforcement of chain energy to sustain FM order of qubits in chains, respectively, lowering
the accuracy of the results measured by D-Wave QA. In addition, even though the D-Wave QA Ocean package
provides a default coupling $J_c^{\text{default}}$, it is not an optimal coupling $J_c^{\text{optimal}}$.
Therefore, determining $J_c^{\text{optimal}}$ is highly desirable for accurate QA measurements.

\begin{figure}
\includegraphics[width=0.95\columnwidth]{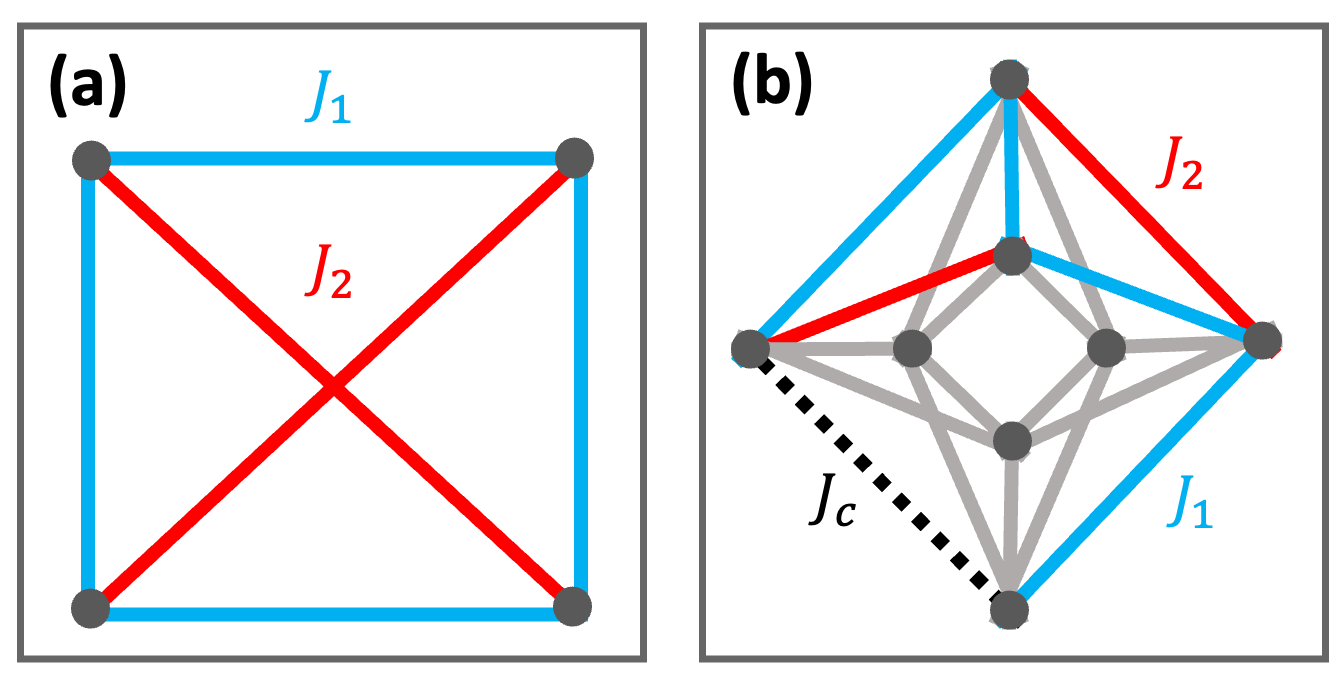}
\caption{\label{Fig1} (Color online) (a) Unit cell of $2 \times 2$ qubits on a two-dimensional (2D) square
structure with the nearest-neighbor coupling $J_1$ and diagonal-neighbor coupling $J_2$ as a combinatorial
optimization problem. (b) Embedding on Pegasus graph for a unit cell of $2 \times 2$
qubits on the 2D square structure. The chains with coupling $J_c$ to sustain ferromagnetic
order of qubits are marked as dotted lines.}
\end{figure}

In this study, we first analyze the full energy spectrum on an embedding of a combinatorial optimization
problem of small size to estimate $J_c^{\text{optimal}}$, through an exact approach that visits
all energy states. We confirm that $J_c^{\text{optimal}}$ would be slightly larger than the
critical chain coupling $J_{\text{c}}^*$ between chain breaking and unbroken state without any chain
breaking. Next, we explain an algorithm by which $J_c^*$ is extracted.
We confirmed that the maximum probability $p^{\text{max}}$ with the possible lowest energy in QA
measurements of various combinatorial optimization problems appears at a chain coupling value that is
$1.2$ times larger than $J_c^*$.

As examples of the combinatorial optimization problems to demonstrate the usefulness of 
$J_c^{\text{optimal}}$ extracted by our algorithm, we
consider a $J_1-J_2$ Ising system on a two-dimensional (2D) $L \times L$ square lattice and
fully connected Ising system with all couplings determined by randomness on $N$ sites.
We present $p$ with the possible lowest energy in many QA shots of both the
combinatorial optimization problems. Finally, it is confirmed that the extracted $J_c^{\text{optimal}}$
showed better $p$ than $J_c^{\text{default}}$ in both the combinatorial optimization problems.

The remainder of this paper is organized as follows. Section~\ref{model} describes the Hamiltonian
made by embedding with chains of FM order between qubits. In Section~\ref{first_cop}, we analyze the full
energy spectrum on the combinatorial optimization problem of the 2D $4 \times 4$ qubits via exact
approach. $9$ qubits are artificially added in the chains. We roughly estimate $J_c^{\text{optimal}}$
from analysis of the exact full energy spectrum result. We also account for the method how
$J_c^{\text{optimal}}$ is extracted. Section~\ref{result} shows that $J_c^{\text{optimal}}$ is better $p$
than $J_c^{\text{default}}$ in QA measurements of $J_1-J_2$ and fully connected combinatorial optimization
problems. Finally, conclusions are presented in Section~\ref{Conclusion}.

\section{Hamiltonian in Embedding\label{model}}

The Hamiltonian of the D-Wave QA is given by
\begin{equation}\label{Eq1}
H = H_{\text{cop}} + H_{\text{chain}},
\end{equation}
where $H_{\text{cop}}$ and $H_{\text{chain}}$ are the parts of the combinatorial optimization problem and
of the chain, respectively. Here, $H_{\text{chain}}$ is expressed as follows:
\begin{equation}\label{Eq2}
H_{\text{chain}} =  -\sum_{i_{\text{chain}}} \sum_{<k,k'>}^{n_{i_{\text{chain}}}}J_{c,i_{\text{chain}}}
\sigma_{i_{\text{chain}},k}^{z} \sigma_{i_{\text{chain}},k'}^{z},
\end{equation}
where $J_{c,i_{\text{chain}}}$ is the coupling of the FM order between $k$ and $k'$ qubits at the
chain site $i_{\text{chain}}$, and $n_{i_{\text{chain}}}$ is the number of qubits. The total number of
qubits $N_{\text{chain}}$ in all chains is given by $N_{\text{chain}}=\sum_{i_{\text{chain}}}
n_{i_{\text{chain}}}$.

\section{Determination of the optimal chain coupling\label{first_cop}}
\subsection{Hamiltonian}

\begin{figure}
\includegraphics[width=0.95\columnwidth]{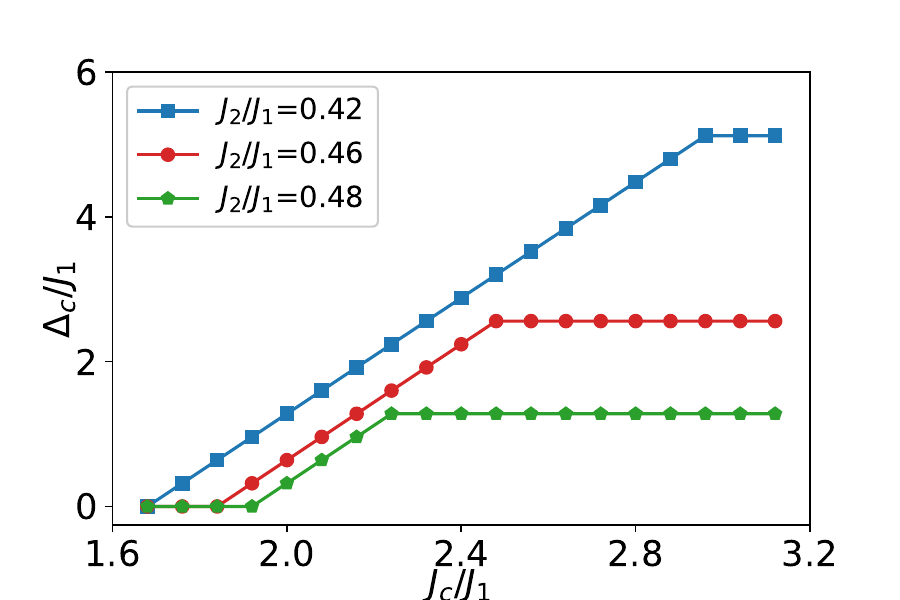}
\caption {\label{Fig2} (Color online) Interval $\Delta_c/J_1$ between the ground and the first
excited energies on the 2D $4 \times 4$ qubits of the combinatorial optimization problem with $9$ qubits
in the chains as a function of $J_c/J_1$ for $J_2/J_1=0.42$, $0.46$, and $0.48$. $\Delta_c/J_1$
are computed by the exact method with visiting all energy states. Note that if any chain is broken, we set
that $\Delta_c/J_1$ is $0$.}
\end{figure}

As an example to show an algorithm how $J_c^{\text{optimal}}$ is determined, we considered a
$J_1-J_2$ combinatorial optimization problem on a 2D $L \times L$ square lattice. The Hamiltonian
$H_{\text{cop}}$ of the $J_1-J_2$ combination optimization problem is defined as follows:
\begin{equation}\label{Eq3}
H_{\text{cop}} = -J_1 \sum_{<i,j>} \sigma_i^z \sigma_j^z + J_2 \sum_{<<i,j'>>} \sigma_i^z \sigma_{j'}^z,
\end{equation}
where the nearest- and diagonal-neighbors are denoted by $<i,j>$ and $<<i,j'>>$,
respectively~\cite{Jin2012,Jin2013}. $J_1$ and $J_2$ mean the nearest-neighbor and diagonal-neighbor
couplings respectively, as shown in Fig.~\ref{Fig1} (a). Here, the reason we chose this $J_1-J_2$
combinatorial optimization problem as the example is because the interval $\Delta$ between the energies of
the ground and of the first-excited states is systematically adjusted by $J_2/J_1$
value~\cite{Jin2012,Jin2013}. $\Delta$ is an important variable that affects the accuracy of QA
measurements, because the plausible local minima close to global minimum in objective function of the
combinatorial optimization problem increase with decreasing $\Delta$. In addition, another energy gap
$\Delta_c$ between the energies of the ground and of the first-excited states in Eq.~(\ref{Eq1})
of an embedded combinatorial optimization problem on Pegasus graph is occurred by chains with
$J_c$. Figs.~\ref{Fig1}(a) and (b) illustrate the unit cell of $2 \times 2$ qubits on the 2D
square structure with the nearest-neighbor coupling $J_1$ and diagonal-neighbor coupling $J_2$ as a
combinatorial optimization problem and its embedding on Pegasus graph, respectively. The chains with the
coupling $J_c$ to keep FM order of qubits are marked as dotted lines in Fig.~\ref{Fig1}(b).

\subsection{Rough estimation of optimal chain coupling by analysis of full energy spectrum\label{exact}}

We analyze the full energy spectrum of $J_1-J_2$ combinatorial optimization problem on a 2D $4 \times 4$
square lattice in the embedding, using the exact method with visiting all energy states. $9$ qubits are
also added in the chains. We find that when all chains are broken in weak $J_c/J_1$, a stable state
occurs. As $J_c/J_1$ increases, the stable state with all chain breakings changes into a state where chain
breaking disappears individually, depending on the chain connecting shape between the qubits of the
combinatorial optimization problem and the chain. This implies that the critical coupling for chain
breaking at each chain depends on the chain connecting shapes. After $J_c^*/J_1$, the state in which any
chain is not broken becomes a stable state.

\begin{figure}
\includegraphics[width=0.95\columnwidth]{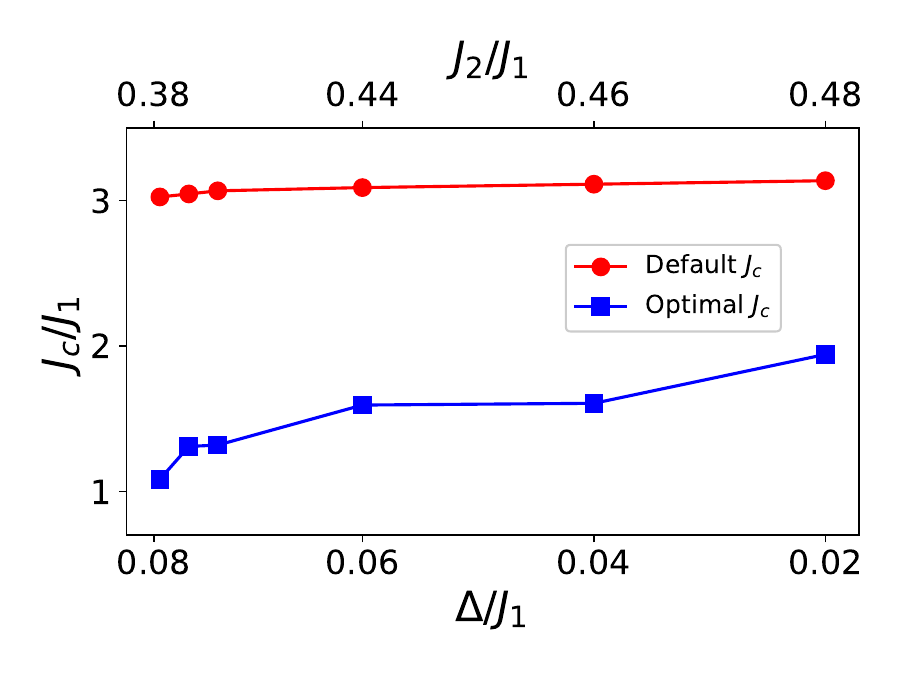}
\caption{\label{Fig3} (Color online) Chain coupling $J_c$ as a function of $\Delta/J_1$ on the bottom
label (and $J_2/J_1$ on top label) in $J_1-J_2$ combinatorial optimization problem on a 2D $8 \times 8$
square lattice. Here, $\Delta/J_1$ in the bottom label are approximately estimated by interval between the
lowest energy and the second lowest energy in many quantum annealing (QA) shots. The top label is $J_2/
J_1$, matching $\Delta/J_1$.}
\end{figure}

Fig.~\ref{Fig2} shows $\Delta_c/J_1$ as a function of $J_c/J_1$ for $J_2/J_1=0.42$, $0.46$, and
$0.48$. Here, we assume that when all chains are unbroken, the most stable state occurs to prevent
deformation of the combinatorial optimization problem. So, if any chain is broken, we set
$\Delta_c/J_1$ to $0$. A stable state without any chain breaking appears above the
critical chain couplings $J_c^*/J_1=1.68$, $1.84$, and $1.92$ for $J_2/J_1=4.2$, $4.6$, and $4.8$,
respectively. $\Delta_c/J_1$ increases as $J_c/J_1$ increases from $J_c^*/J_1$ until kink behavior is
observed at $J_c^{**}/J_1=2.96$, $2.48$, and $2.24$ for $J_2/J_1=4.2$, $4.6$, and $4.8$, respectively.
After $J_c^{**}/J_1$ for all $J_2/J_1$, $\Delta_c/J_1$ converges to $\Delta/J_1$. We believe that
the chain breaking and dominant chain energy states that sustain the FM order of the qubits in the chain
would appear below $J_c^*/J_1$ and above $J_c^{**}/J_1$, respectively. The range between $J^*$ and
$J^{**}$ decreases with increasing $J_2/J_1$ (or decreasing $\Delta/J_1$). Finally, we inferred three
possibilities from exact result in Fig.~\ref{Fig2}: (i) $J_c^{\text{optimal}}/J_1$
might be between $J_c^*/J_1$ and $J_c^{**}/J_1$. (ii) When $\Delta/J_1$ is tiny, it may be difficult to
determine $J_c^{\text{optimal}}/J_1$. (iii) The accuracy of QA measurements will be poor with decreasing
$\Delta/J_1$.

\subsection{Procedure for extracting the optimal chain coupling $J_c$ value\label{exact}}

The D-Wave Ocean package provides the default coupling $J_c^{\text{default}}$ given as
\begin{equation}\label{Eq4}
J_c^{\text{default}} = 1.41 * (\frac{1}{M_c} \sum_{<i,j>} J_{ij}^2)^{\frac{1}{2}} * (\frac{1}{N_c} \sum_{i=1} \text{deg}(i))^{\frac{1}{2}},
\end{equation}
where $M_c$, $N_c$ and $\text{deg}(i)$ are the number of total connectivities between qubits, the number
of total qubits and the coordination number at $i$-chain, respectively. Even though $J_c^{\text{default}}$
gives not bad results in accuracy of QA measurements, we confirmed that
$J_c^{\text{default}}$ is not $J_c^{\text{optimal}}$ from estimation of the
exact result in Fig.~\ref{Fig2} and from various results measured by D-Wave QA.

Here, we design more sophisticated algorithm to obtain $J_{\text{c}}^{\text{optimal}}$ that provides
better results than $J_c^{\text{default}}$. The procedure of the algorithm is followed as: (i) In the
first step, we set to initial chain coupling $J_c$ given as $J_c=0.1 \times J_c^{\text{default}}$ of weak
chain coupling. The stable state at initial $J_c$ is when almost all chains are broken. (ii) In the second
step, we increase $J_c$ by $10$ percent of $J_c^{\text{default}}$ and perform QA measurements. The
chain breaking gradually disappears with increasing $J_c$. (iii) In the third step, we set the critical
chain coupling $J_c^*$, when the chain is no longer broken. (iv) Finally, we set $J_c^*$ multiplied by
$1.2$ as $J_c^{\text{optimal}}$. The factor $1.2$ is obtained empirically from the best results of QA
measurements in various combinatorial optimization problems on D-Wave Advantage machine. In open code we
set annealing time to $1.0 \times 10^{-3}$ second in QA measurements, and perform a hundred of annealing
shots to find $J_c^*$ and $J_c^{\text{optimal}}$.

\begin{figure}
\includegraphics[width=0.95\columnwidth]{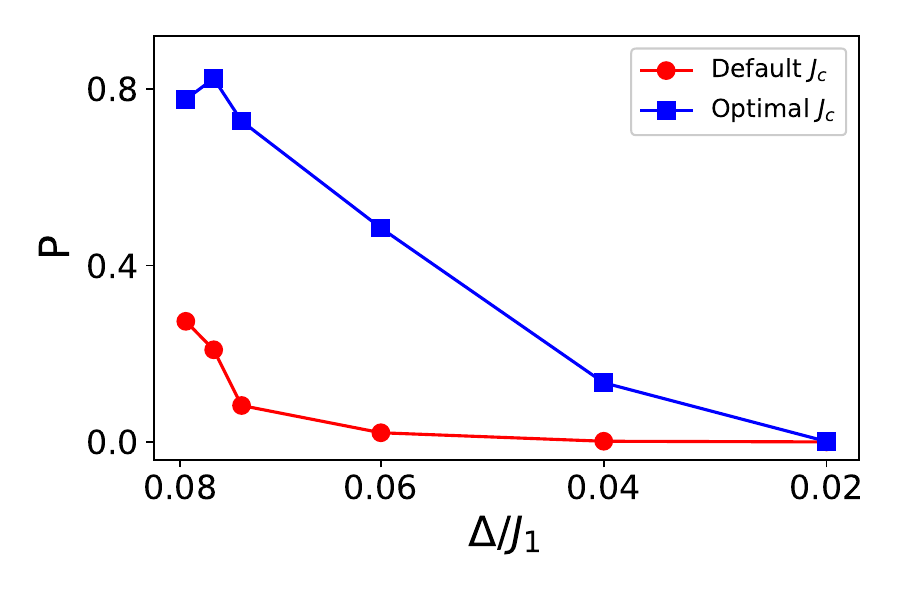}
\caption{\label{Fig4} (Color online) Probability $p$ for observing the possible lowest energies measured
by $5000$ times QA shots with $J_c^{\text{default}}$ and $J_c^{\text{optimal}}$ on $J_1-J_2$ combinatorial
optimization problem on 2D $8 \times 8$ square lattice.}
\end{figure}

\section{results\label{result}}
\subsection{$J_1-J_2$ combinatorial optimization problem\label{exact}}
In order to demonstrate the superiority of the results in $J_c^{\text{optimal}}$ extracted by our
algorithm, we first consider $J_1-J_2$ combinatorial optimization problem on 2D $8 \times 8$ square
lattice expressed by Hamiltonian of Eq.~(\ref{Eq3}). We compare $J_c^{\text{default}}/J_1$ computed by
Eq.~(\ref{Eq4}) to $J_c^{\text{optimal}}/J_1$ determined by our algorithm. Fig.~\ref{Fig3} shows
$J_c^{\text{default}}$ and $J_c^{\text{optimal}}$ values as a function
of $\Delta/J_1$ in bottom label (and $J_2/J_1$ in top label). Here, $\Delta/J_1$ values in bottom label
are approximately estimated by interval between the lowest energy and the second lowest energy in many QA
shots. The top label is $J_2/J_1$, matching $\Delta/J_1$. We observe that $J_c^{\text{default}}/J_1$
values are larger than $J_c^{\text{optimal}}/J_1$ in all $\Delta/J_1$. $J_c^{\text{default}}/J_1$
increases slowly, while $J_c^{\text{optimal}}/J_1$ increases quickly as $\Delta/J_1$ decreases.

In addition, we plot $p$ for observing the possible lowest energies measured by QA shots of $5000$
times with both $J_c^{\text{default}}/J_1$ and
$J_c^{\text{optimal}}/J_1$ in Fig.~\ref{Fig4}. $p$ (or accuracy of QA measurements) decrease with
decreasing $\Delta/J_1$ in both $J_c^{\text{default}}/J_1$ and $J_c^{\text{optimal}}/J_1$, because many
plausible local minimum that are similar to the global minimum are created in tiny $\Delta/J_1$. As a main
result, we find that $p$ values measured in $J_c^{\text{optimal}}/J_1$ are higher than those in
$J_c^{\text{default}}/J_1$. We think that $J_c^{\text{optimal}}/J_1$ provides the optimal compromise
between chain breaking and chain energy, while $J_c^{\text{default}}/J_1$ lowers the accuracy of QA
measurements by the strong chain energy.

\begin{figure}
\includegraphics[width=0.95\columnwidth]{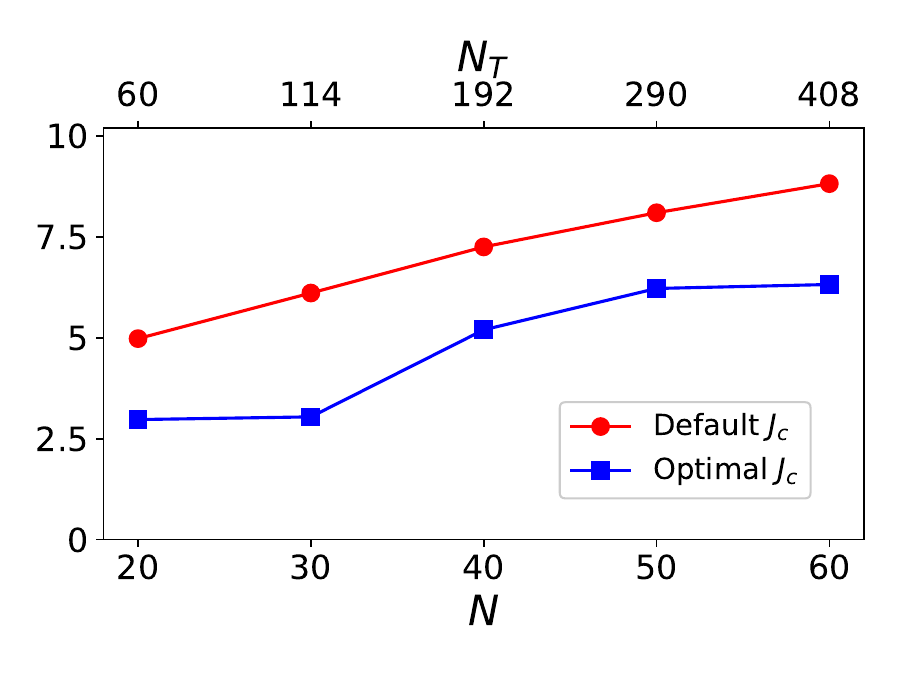}
\caption{\label{Fig5} (Color online) $J_c$ as a function of $N$ on bottom label (and $N_T$ on top label)
in fully connected combinatorial optimization problem. Here, $N$ on bottom label and $N_T$ on top label
are the number of qubits in fully connected combinatorial optimization problem and the number of qubits in
embedding, respectively.}
\end{figure}

\subsection{Fully connected combinatorial optimization problem\label{exact}}
Finally, in order to confirm a more general case we select a fully connected combinatorial optimization
problem, where obtaining accurate results of QA measurements is more difficult than $J_1-J_2$
combinatorial optimization problem because it produces more plausible local minima than $J_1-J_2$
combinatorial optimization problem, due to strong degeneracy in energy states.
The Hamiltonian $H_{\text{cop}}$ is given as
\begin{equation}\label{Eq5}
H_{\text{cop}} = -\sum_{<i,j>}^{N} J_{<i,j>} \sigma_i^z \sigma_j^z,
\end{equation}
where $<i,j>$ and $N$ mean all couplings between qubits and the number of qubits in the combination
optimization problem, respectively. $J_{<i,j>}$ values are either $-1.0$, $0.0$ or $1.0$ determined by
random number generator.

\begin{figure}
\includegraphics[width=0.95\columnwidth]{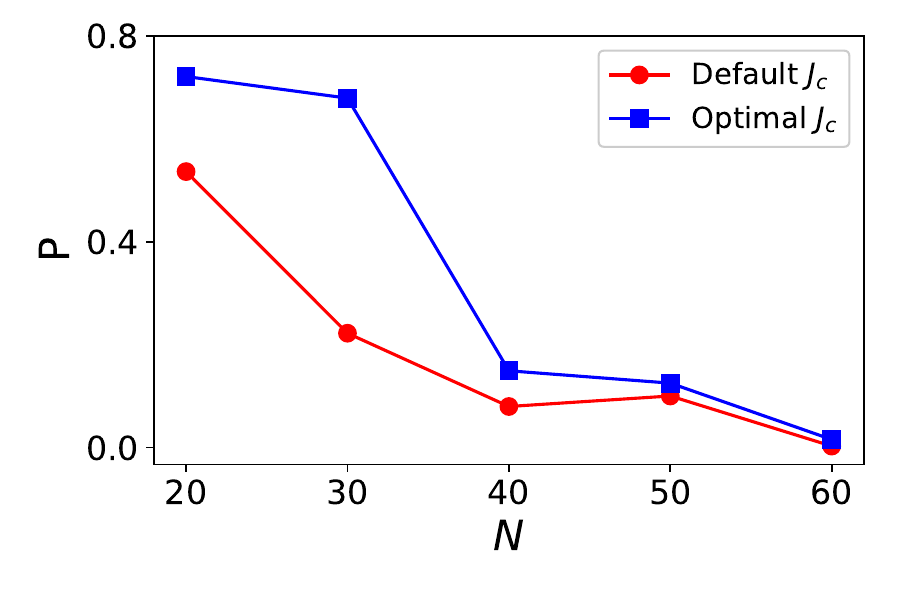}
\caption{\label{Fig6} (Color online) $J_c$ as a function of $N$ in fully connected combinatorial
optimization problem. $p$ are measured by QA shots of $5000$ times with $J_c^{\text{default}}$ and
$J_c^{\text{optimal}}$.}
\end{figure}

Same as $J_1-J_2$ combinatorial optimization problem, we plot $J_c$ values and $p$ as a function
of $N$ on bottom label in Fig.~\ref{Fig5} and Fig.~\ref{Fig6}, respectively. $p$ are measured by QA shots 
of $5000$
times with $J_c^{\text{default}}$ and $J_c^{\text{optimal}}$. $J_c^{\text{default}}$ values are
always bigger than $J_c^{\text{optimal}}$ values in all $N$ of Fig.~\ref{Fig5}. The number of total qubits
$N_T$ on top label in embedding is increasing with $N$ in Fig~\ref{Fig5}. It is also confirmed
that $J_c^{\text{optimal}}$ gives more accurate results than $J_c^{\text{default}}$ in QA measurements
of Fig.~\ref{Fig6}.

\section{Conclusion\label{Conclusion}}

The D-Wave QA displays significant potential due to the recent rapid increase in qubit capacity.
Moreover, the D-Wave QA has been extensively applied to solving combinatorial optimization problems.
However, as a limitation, physical embedding with FM chains of several qubits is required to consider the
exact long-range coupling between qubits ignored in D-Wave QA. The separation of FM qubits in the chains
or the dominant enhancement of the chain energy to sustain FM order of qubits in chains lowers the
accuracy of the results measured by D-Wave QA. In addition, we confirm that even though the D-Wave
Ocean package provides the default coupling $J_c^{\text{default}}$, it is not
an optimal coupling $J_c^{\text{optimal}}$. Therefore, determining the optimal coupling
$J_c^{\text{optimal}}$ for maintaining the appropriate FM ordered qubits in the chains is required to
improve accuracy.

In this paper, we present the algorithm how $J_c^{\text{optimal}}$ is determined. We confirm that the
extracted $J_c^{\text{optimal}}$ shows much better $p$ than $J_c^{\text{default}}$ in various parameters
of $J_1-J_2$ and fully connected combinatorial optimization problems. Finally, We believe that our method
provides practical assistance for D-Wave QA users to search for $J_c^{\text{optimal}}$ with decreased
essential errors. The open code is available in \textit{https://github.com/HunpyoLee/
OptimizeChainStrength}.

\section{Acknowledgments}
This work was supported by Institute of Information and communications Technology Planning Evaluation 
(IITP) grant funded by the Korean government (MSIT) (No. RS-2023-0022952422282052750001), Ministry of
Science through NRF-2021R1111A2057259 and by the quantum computing technology development program of the
National Research Foundation of Korea (NRF) funded by the Korean government (Ministry of Science and
ICT(MSIT)) (No. 2020M3H3A1110365).

\section{Data Availability\label{data}}
The data that support the findings of this study are available from the corresponding author upon 
reasonable request.

\section{Conflict of interest\label{interest}}
The authors have no conflicts to disclose.

\bibliography{paper}

\end{document}